# Network dynamics of community resilience and recovery: new frontier in disaster research

Chia-Fu Liu[1]✉, Ali Mostafavi[1]

Disasters impact communities through interconnected social, spatial, and physical networks. Analyzing network dynamics is crucial for understanding resilience and recovery. We highlight six studies demonstrating how hazards and recovery processes spread through these networks, revealing key phenomena, such as flood exposure, emergent social cohesion, and critical recovery multipliers. This network-centric approach can uncover vulnerabilities, inform interventions, and advance equitable resilience strategies in the face of escalating risks.

## 1      Introduction

The central argument of this perspective paper is that community resilience and recovery cannot be fully understood or effectively strengthened without addressing the underlying network dynamics—encompassing the social, spatial, and physical structures embedded in communities. Disasters propagate through these interconnected webs, extending their impacts far beyond immediate hazard zones and shaping the profiles of subpopulations that would receive information, resources, and support[1,2]. At the same time, recovery does not occur in isolation: it relies on the diffusion of assistance, collective sense-making, and interdependent processes that unfold within these networks. Importantly, network structures also influence inequalities in disasters[3], as those with fewer connections or resourceful ties experience slower and less robust recoveries. By leveraging emerging data and novel modeling techniques, we can capture these hidden socio-spatial interdependencies, identify high-impact "recovery multipliers[4]," and develop targeted interventions that catalyze community-wide resilience. Essentially, network science offers a new frontier for both theoretical and practical advancements in disaster research[5].

During the past two decades, scholarship on network science has illuminated how the topology and dynamics of connections - among individuals, organizations, neighborhoods or even larger infrastructure systems - govern complex phenomena[6,7]. In fields such as epidemiology[8], it has long been understood that infection patterns emerge from social interaction networks rather than from a simplistic notion of population averages. Similarly, in communication studies, the spread of information (and misinformation) depends on who is connected to whom, how often, and through what channels[9,10]. More recently, disaster research has started to integrate network-based approaches, suggesting that community resilience and recovery operate through multiple overlapping networks of social ties, physical infrastructure, and spatial mobility flows[11–13]. Viewed through this lens, communities comprise interdependent systems of people, resources, and built environments. These systems are not static; they evolve with urban development, demographic changes, technology adoption, and broader socio-economic forces. Disasters can transform network structures by disrupting roads, severing supply chains, or fracturing social groups. Conversely, strong network connections—or the ability to rapidly reconfigure them—can facilitate the flow of information, resources, and support in ways that promote timely and equitable recovery. Thus, a network perspective promises to reveal hidden dynamics that shape outcomes in both acute crises and long-term rebuilding efforts.

[1]Department of Civil and Environmental Engineering Texas A&M University, College Station, TX 77843, USA. ✉e-mail: joeyliu0324@tamu.edu

Network dynamics research moves beyond static snapshots of who is connected to whom, illuminating how and why interactions and influence propagate over time across interconnected systems[14–16]. In complex domains such as community resilience, merely mapping a network's structure or topology provides an incomplete view: it tells us where potential linkages exist but not how a disturbance (e.g., a hazard, resource shortage, or cascade failure) traverses those linkages in real-world contexts. Dynamic analysis captures temporal processes—including feedback loops, adaptive behaviors, and threshold effects—that drive transformative changes in networked systems[17,18]. By focusing on how networks "behave" under stress, we reveal emergent properties, such as collective sense-making[19], recovery multipliers[4,20], or cascading failures[21], each of which depends on the timing and sequence of interactions rather than just the overall connectivity pattern. A core premise of network dynamics-based disaster research is that networks are the structures upon which resilience and recovery processes unfold. Whether these networks are social (ties between individuals, families, organizations), spatial (patterns of human mobility, commuting flows, or neighborhood adjacencies), or physical (utility grids, transportation systems), they collectively enable or constrain how a community responds to a shock. The connectivity of these networks determines how rapidly assistance can arrive, how soon critical infrastructure can be restored, and how effectively community members can coordinate to meet shared needs[22,23]. In this sense, resilience is not just an inherent property of an individual, a building, or a neighborhood, rather it is an emerging attribute of interdependent systems. Each node (be a person, a business, or an infrastructural asset) is influenced by what happens to neighboring nodes. For instance, a business may be able to reopen its doors quickly, but if nearby suppliers or complementary businesses remain shut, customer flows could remain suppressed, slowing the overall rebound. Alternatively, an influential social media user might rapidly disseminate updates on relief resources, facilitating broader community awareness and uptake. In both cases, the network position of the entity—how central or peripheral, how many connections it has—critically shapes collective outcomes.

One of the most significant catalysts propelling network-based disaster research is the availability of high-resolution data on social, spatial, and economic interactions. The widespread adoption of smartphones, geolocation services, and digital platforms (from social media to ride-sharing applications) has generated massive, fine-grained datasets that can reveal mobility patterns, communication structures, and resource flows in near-real-time. For instance, anonymized GPS traces can show how people move before, during, and after a flood, allowing analysts to identify both highly exposed routes and hubs of relative safety[24–26]. Social media content can illuminate how risk perceptions and coping strategies spread across online communities[19,27]. These emerging data enable computational modeling approaches—such as agent-based simulations[25], threshold-based diffusion models[4,20], and machine learning[28] for pattern detection—that were scarcely feasible two decades ago. Researchers can now create dynamic representations of socio-spatial networks, where each node and edge is characterized by varying degrees of hazard exposure, connectivity, and resilience capacity. Such models can capture complex feedback loops: for example, how infrastructure failure in one location propagates to other areas, or how misinformation about the disaster modifies evacuation behavior. Furthermore, these datasets allow for empirical testing of theories on community resilience, bridging the gap between conceptual frameworks and observable metrics.

Several cross-cutting themes have emerged from recent explorations into the network dynamics of disaster resilience. First is the notion of diffusion—the process by which impacts (or recovery) spread through interconnected nodes[4,24]. Diffusion might refer to the spread of hazard exposure (e.g., viruses, floodwaters) or the propagation of beneficial resources (e.g., supplies, information, capital). Second, inequality is a recurring concern, as network structures often map onto socio-economic divides, reinforcing disparities in how quickly and comprehensively different groups recover[29]. Third, critical nodes—variously described as hubs, multipliers, or bridges"—play crucial roles in shaping collective outcomes[4,20]. Identifying and activating these nodes can be a powerful strategy to accelerate recovery and bridge resource gaps. In the following sections, we illustrate these ideas through six studies that represent diverse yet complementary perspectives on network-based disaster research. As shown in Figure 1, these studies collectively underscore the multifaceted ways in which social, physical, and spatial networks shape how communities experience, respond to, and ultimately recover from disruptive events.

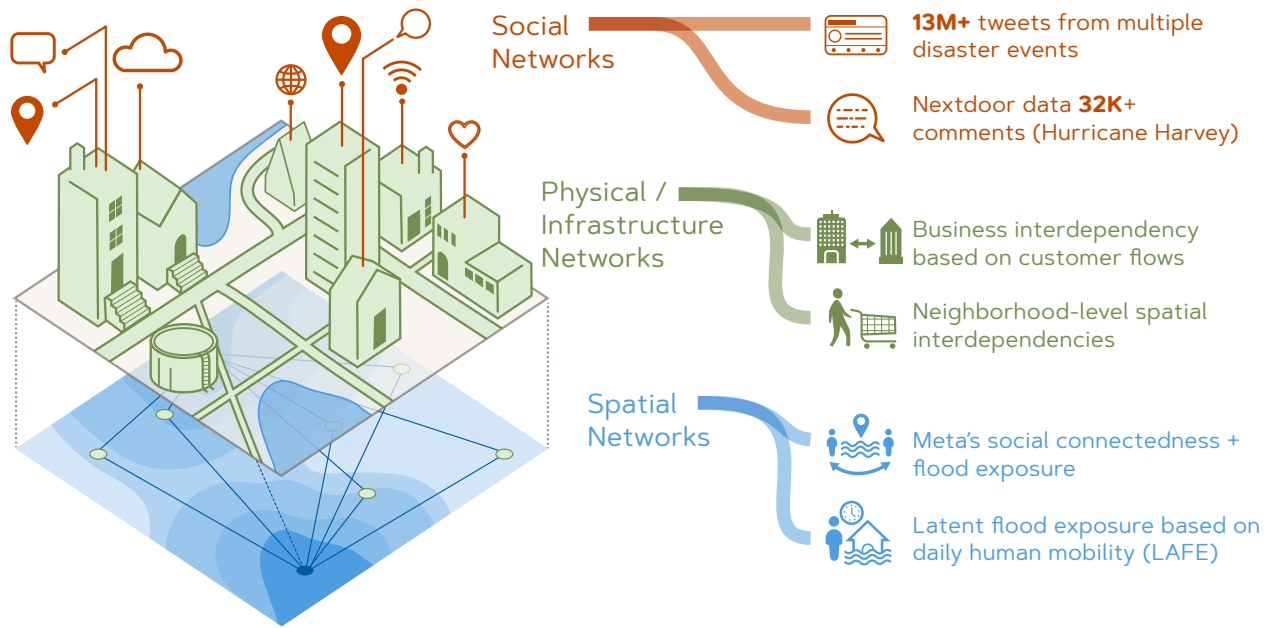

**Figure 1 | Conceptual illustration.** The six studies explore different aspects of urban networks—social, physical/infrastructure, and spatial—to characterize community resilience and recovery.

## 2      Network dynamics processes in disasters

This section presents six studies, each examining different aspects of network dynamics in disasters. By addressing the social, spatial, and physical aspects of urban networks, they collectively underscore our central contention that network structures and processes are pivotal to understanding—and ultimately improving—community resilience and recovery. Table 1 summarizes these studies, highlighting their network focus and dynamics.

**Table 1 | Summary of six key studies on network dynamics in disaster research**

| Study & reference | Network focus | Key network dynamics concept | Data/approach | Key findings | Relevance to resilience | Novel concepts & characterizations |
|---|---|---|---|---|---|---|
| Dynamics of collective information processing for risk encoding in social networks during crises[19] | Social media networks (*Twitter*) | • Stability of network structures despite new users<br>• Power-law influence distribution<br>• Spatial clustering of risk discussions | • Analysis of 13M+ tweets from multiple disaster events<br>• Network measures of posting, retweeting, quoting, etc. | • A small subset of influential communicators shapes information flow<br>• Despite global reach, risk discussions remain localized<br>• Network structures remain consistent over time | • Highlights influential nodes as crucial for risk communication<br>• Shows information diffusion is not uniform but funneled through hubs<br>• Reinforces that collective sense-making relies on both social hierarchy and geographic proximity | • Influential hubs in crisis diffusion (power-law dynamics)<br>• Temporal invariance of network structure during disasters<br>• Conceptual emphasis on persistent local clustering in risk discourse |
| Dynamics of post-disaster recovery in behavior-dependent business networks[20] | Business interdependency networks | • Recovery as a diffusion process among interdependent businesses<br>• "Recovery multipliers"<br>• Socioeconomic disparities in recovery roles | • Network diffusion model based on behavioral dependencies (customer flows)<br>• Genetic algorithm optimization for identifying multiplier businesses | • Certain businesses, if they recover early, accelerate the recovery of others<br>• "Recovery blockers" impede broader network rebound<br>• Different businesses serve as multipliers in high- vs. low-income areas | • Advocates targeted interventions at key businesses to amplify local economic recovery<br>• Shows how inequities shape which nodes accelerate or slow overall recovery | • Recovery multipliers (key businesses accelerating wider rebound)<br>• Recovery blockers limiting network-level restoration<br>• Formalizing behavior-based interdependencies in economic resilience |
| Human-centric characterization of life activity flood exposure shifts focus from places to people[24] | Spatial mobility networks | • Measuring flood exposure based on daily movement (LAFE)<br>• "Latent exposure" vs. "latent immunity"<br>• Influence of urban form on population risk | • Anonymized smartphone mobility data across 18 U.S. cities<br>• Time-based dwell analysis in flood zones | • Traditional flood maps can under- or overestimate actual risk<br>• Higher road density correlates with greater exposure<br>• Polycentric cities disperse risk more effectively | • Demonstrates mobility-driven flood risk, not just residence-based<br>• Suggests infrastructure and urban planning can reduce real-world flood exposure<br>• Underscores dynamic network assessment of risk | • Life Activity Flood Exposure (LAFE) as a new metric for risk<br>• Latent hazard exposure across daily movement<br>• Concept of latent immunity when exposed populations spend time outside flood zones |
| Revealing hazard-exposure heterophily as a latent characteristic of community resilience in social-spatial networks[29] | Socio-spatial networks | • Hazard-exposure heterophily vs. homophily<br>• Resourceful vs. non-resourceful ties<br>• Income segregation as network isolation | • Meta's Social Connectedness Index (SCI) + flood data in Harris County, Texas<br>• Measuring cross-hazard social ties | • Communities linking high- and low-exposure zones recover faster<br>• Low-income areas have stronger hazard-exposure homophily, limiting external help<br>• Social connectivity across flood boundaries is a key resilience factor | • Identifies social ties across hazard boundaries as crucial for resource exchange<br>• Illuminates how income segregation exacerbates vulnerability<br>• Suggests targeted strategies to increase heterophily for more equitable recovery | • Hazard-exposure heterophily reflecting cross-boundary resilience ties<br>• Resourceful vs. non-resourceful social connections<br>• Emphasis on income segregation as a driver of network isolation |
| Network diffusion model reveals recovery multipliers and heterogeneous spatial effects in post-disaster community recovery[4] | Neighborhood-level spatial interdependencies | • Threshold-based spatial diffusion of recovery<br>• "Recovery multipliers" at neighborhood scale<br>• Importance of coordinated interventions | • Hurricane Harvey (2017) data for Houston, Texas<br>• Census block group–level location-based activity as proxy for recovery<br>• Threshold-based diffusion modeling | • Neighborhoods vary in speed of recovery diffusion<br>• Low-income/minority communities often serve as critical recovery multipliers<br>• Spatial adjacency fosters or inhibits community-wide recovery | • Shows coordinated interventions in key neighborhoods produce system-wide benefits<br>• Reinforces that recovery spreads (or stalls) through local connectivity<br>• Demonstrates the power of data-driven spatial diffusion models for planning | • Recovery multipliers (community-level) & recovery isolates<br>• Heterogeneous recovery thresholds influencing diffusion<br>• Formalization of spatial adjacency in cascading recovery |
| Emergent social cohesion for coping with community disruptions in disasters[27] | Online social networks (*Nextdoor*) | • Emergent social cohesion through abrupt, extensive tie formation<br>• Role of weak ties in crisis information flow<br>• Diminished role of geographic homophily | • Hurricane Harvey (2017) data from 28 Houston neighborhoods<br>• 2,690 active users, 1,939 posts, 32,776 comments<br>• Four NRT modalities: enactment, activation, reticulation, performance | • Online activity spikes alongside disaster onset<br>• Thematic discussions shift from damage/shelter to volunteer/relief<br>• Weak ties bridge neighborhoods, enabling rapid information dissemination<br>• Network assortativity evolves as crisis subsides | • Highlights temporary yet impactful social ties that enhance collective resilience<br>• Shows neighborhood-based online platforms facilitate cross-boundary collaboration<br>• Underscores how emergent connectivity can strengthen community response | • Emergent social cohesion driven by crisis conditions<br>• Weak ties as critical bridges for widespread communication<br>• Network reticulation theory linking abrupt tie formation to disaster triggers |

## 2.1 Latent hazard exposure shaped by spatial mobility networks

Disasters, though localized in origin, often trigger cascading effects that extend far beyond the immediate impacted zone. Traditional hazard exposure assessments commonly rely on place-based approaches that compare hazard data with residential population distributions[30,31]. Although useful, these methods neglect a critical dimension of human behavior: daily activity patterns. People's lives extend beyond residential locations to workplaces, schools, recreation areas, and shopping centers. Consequently, a substantial portion of hazard exposure occurs outside residential areas, presenting a blind spot in conventional evaluations.

To address this gap, a recent study introduced the concept of the Life Activity Flood Exposure Framework (LAFE), which uses human life activity networks to evaluate flood exposure in urban settings[24]. Instead of relying solely on static maps that label certain areas as flood zones and assume residents within those zones are most exposed, the study measures how much time individuals actually spend in flood-prone areas based on anonymized smartphone mobility data. By tracking real movement patterns, researchers developed the LAFE metric, revealing that some people living outside hazard zones experience high exposure due to regular travel into risky areas, while others living inside flood zones may have latent immunity if they spend most of their day elsewhere. This person-centered approach highlights how urban form and travel networks shape the actual flood risk. Two cities with similar flood extents may see starkly different LAFE distributions depending on commuting patterns, the location of workplaces, and how retail or leisure facilities are distributed. In addition, the study notes that changes in road networks or public transit can exacerbate or alleviate these risk patterns over time.

This research underscores the spatial dimension of disaster resilience in the network, showing that exposure to hazards is not merely a matter of geographic 'in' or 'out', but a function of mobility-driven connectivity. The implication is that flood impacts (and possibly other hazards) can extend beyond their presumed boundaries through movement patterns in daily life. Such findings reinforce the core notion that network dynamics—in this case, commuting flows and daily travel routes—are essential to understanding who is truly at risk. By shifting the focus from static place-based metrics to behavior-based metrics, the study exemplifies how innovative data and models can reveal hidden vulnerabilities. This approach aligns with the overarching argument that resilience strategies must go beyond place-based zoning to incorporate the human activities that shape real-world exposure and recovery potentials.

### 2.2 Hazard Exposure Heterophily within Socio-spatial Networks

A key element of resilience lies in social capital, which strengthens community networks. For example, social networks can act as channels for mobilizing relief resources, such as information sharing or physical support, during hazards[32,33]. These invisible ties connect communities, functioning as critical infrastructure during crises. By mapping these networks, researchers can identify gaps and opportunities to bridge connections between affected and unaffected areas, enhancing the effectiveness of disaster response and resource allocation. Social ties manifest in various ways, with one of the most explicit being friendship connections on social media. A recent study[29] addresses a different but equally critical dimension of socio-spatial connectivity: who is socially connected to whom across hazard boundaries. Introducing the concept of hazard-exposure heterophily, the study shows that communities are more resilient when people in flood-prone areas maintain social ties with people in safer zones. These "resourceful" ties can offer critical external support, such as financial help, temporary shelter, or emotional assistance, thereby speeding up recovery. Conversely, hazard-exposure homophily—the situation in which at-risk individuals are primarily connected with others who are similarly exposed—can trap communities in a vulnerability feedback loop.

Before this integration, the literature often linked slower recovery and greater impacts in socioeconomically disadvantaged areas to inherent socioeconomic characteristics. This study, however, reveals another critical factor: the socio-spatial network embedded within communities. For instance, socially vulnerable populations in high-risk hazard zones tend to connect with similarly exposed communities. This "hazard exposure homophily" may explain the prolonged recovery periods in such areas, as it limits access to resourceful connections that could expedite recovery efforts. By juxtaposing community socio-spatial networks with hazard exposure characteristics, this research highlights the need to incorporate underlying network structure into resilience strategies. In addition, the research finds that income segregation is strongly correlated with hazard-exposure homophily. Lower-income areas situated in high-risk zones tend to be socially isolated from resource-rich communities, further inhibiting their post-disaster rebound. The authors suggest that measuring resourceful-tie rates can serve as a diagnostic tool for identifying which areas might require targeted outreach or policy interventions.

This study directly connects to the discussion of network structures and their role in exacerbating or mitigating inequality. It offers a powerful demonstration of how social ties can be more or less "resourceful" depending on whether they cross hazard boundaries. The emphasis on heterophily resonates with the idea that bridging networks—ties linking diverse groups—fosters resilience by providing flows of aid and information. This underscores our argument that resilience is an emergent network property: the presence or absence of cross-boundary links can significantly shape a community's capacity to recover. Moreover, it suggests that interventions aiming to increase hazard-exposure heterophily (e.g., through housing policies, digital connectivity programs, or community-building initiatives) may be critical for addressing socio-spatial disparities.

**2.3   Diffusion in business networks and economic resilience**

A recent study of the *Dynamics of Post-disaster Recovery in Behavior-dependent Business Networks*[20] shifts attention to economic recovery networks. Focusing on the aftermath of Hurricane Ida (2021) on the Louisiana Gulf Coast, the study argues that business recovery unfolds through behavior-based dependencies; for example, how customer mobility and visitation patterns create interlocking fates among neighboring businesses. By applying a network diffusion model, they show that certain establishments, called "recovery multipliers," if they reopen quickly, can act as catalysts that speed up recovery for the wider network.

In addition, the study highlights socioeconomic disparities in recovery. The critical "multiplier" businesses differ by neighborhood income level: in higher-income areas, retail or wholesale businesses often serve as anchors of economic revival, while in lower-income areas, essential service providers (e.g., auto repair, personal care) play that critical role. This implies that uniform policy measures—like blanket financial assistance—may yield suboptimal results if not tailored to the specific network structures and business interdependencies of an affected locality.

The notion of business interdependence strongly parallels the broader concept of network dynamics in community resilience. Just as a disease can spread among connected individuals, or information can diffuse through social media, business revival can propagate across a network shaped by daily consumer behavior. This study makes explicit the concept of recovery multipliers, reinforcing our central argument that certain nodes in a network are especially influential and warrant targeted intervention. By identifying and supporting such nodes, policymakers can effectively activate beneficial diffusion processes, leading to faster and more equitable recovery across the entire community. This approach resonates with the perspective that resilience interventions should harness existing network pathways rather than tackling each node in isolation.

**2.4    Risk information diffusion and collective sense-making**

One remarkable aspect of human society is its ability to process and share information. Within communication networks, knowledge spreads through exchange between agents, resulting in dynamic network diffusion. This process, visible across various network types, encompasses cascading behaviors, the spread of ideas, information, influence, and diseases. These phenomena offer critical insights into community response during disasters and recovery post-disasters. Studying network diffusion involves two primary challenges: (1) tracking the cascading processes—observing how influence flows within the network over time, and (2) identifying diffusion mechanisms—pinpointing sources of influence and understanding their impact.

Two studies leverage crisis informatics diffusion mechanisms to explore collective information processing during disasters using online social media data[19,27]. Social media networks serve as proxies for studying the flow of crisis-related information and the influence of social ties during disasters. Addressing the first challenge, online social media data capture users' communication behavior, making it possible to trace cascading processes. The diffusion mechanism can be summarized as follows: when a disruptive event occurs, social media users post, share, and respond to the event to disseminate impact information and adjustment responses. By analyzing the structure and evolution of these communication networks, these studies reveal the interplay between disruptive events, user activities, and the transformation of network structures. This dynamic process enhances information propagation and collective action, emphasizing the critical role of social ties in fostering situational awareness and coordinated responses. One study[19] delves into the proportions of various communication activities—posting, reposting, replying, and quoting messages—on social media during crises. Key findings indicate that crisis events trigger communication spikes, enlarge social networks, and increase user participation in information sharing. Most users prefer reposting content over generating original messages, demonstrating a tendency to consume and disseminate information to enhance situational awareness before making critical decisions, such as evacuation. In addition, activity proportions across hazard events and spatial contexts remain consistent, indicating universal patterns in information-sharing behavior. The study also reveals that a small group of influential communicators drives the majority (99%) of information propagation. However, these communicators receive limited feedback from information consumers. These findings underscore the importance of collaborating with influential communicators to amplify messaging. Relief organizations and government agencies, not inherently influential on social media, could engage these communicators to boost crisis response effectiveness.

**2.5    Emergent social cohesion enhances risk encoding**

Another study[27] investigates emergent social cohesion as a dynamic network process in online social networks during disasters. This study illuminates how emergent social cohesion arises in disaster contexts through neighborhood-based social media by focusing on Nextdoor communications during Hurricane Harvey (2017). By applying network reticulation theory (NRT), the authors analyze how sudden extensive ties form among residents who previously lacked strong connections, highlighting weak social ties—rather than preexisting relationships—as critical conduits for rapid information flow. Using data from 28 Houston neighborhoods (2,690 users, 1,939 posts, 32,776 comments), researchers tracked daily interactions and topics across four modalities: enactment (disruptive events triggering social interaction), activation (thematic clusters of discussion), reticulation (formation of new ties), and performance (structural stability of emergent networks). Their findings reveal that online activity peaks and shifts alongside the hurricane's phases, with conversation themes moving from infrastructure damage and shelter need to volunteer coordination. Notably, cross-neighborhood communication reduces geographic homophily, while high-degree users frequently link with lower-degree nodes, indicating disassortative mixing that facilitates broad information dissemination. Over time, the network becomes more assortative among similarly active participants, reflecting a post-crisis stabilization of newly formed ties. Overall, this study underscores how weak ties, amplified by neighborhood-tagged social platforms, catalyze shared awareness and resource sharing, illustrating the role of dynamic emergent connectivity in strengthening community resilience during crises.

These studies demonstrate how communities use online social networks (particularly Twitter) to make sense of risk during disasters. By analyzing millions of tweets across multiple disaster events (hurricanes, wildfires, power outages), the findings reveal a paradox of social media networks: despite the influx of new users and conversations, network structures remain remarkably stable. Communication patterns maintain a hierarchical shape dominated by a relatively small set of influential users who disseminate the majority of risk information. Moreover, the studies find that spatial proximity still matters in digital interactions: risk discussions are highly localized, with communication intensity decreasing for users farther from the hazard. This result challenges the assumption that social media platforms automatically remove geographic barriers. Instead, risk encoding (i.e., how communities interpret and label hazard-related information) follows a network pattern that is both socially and spatially constrained. These studies exemplify the diffusion process in a social network context: risk knowledge does not spread evenly but rather is funneled through influential "hub" accounts. By identifying such hubs, emergency managers and policymakers could target these communicators with verified updates, reducing misinformation and accelerating clarity during disasters. Overall, the study underscores how social network structure—rather than just the hazard's physical intensity—plays a determining role in shaping collective sense-making and, by extension, community resilience.

## 2.6   Post-disaster recovery interdependency among neighborhoods

Another recent study of mobility-mediated post-disaster recovery[20] highlights the importance of network effects in community recovery. This study focuses on recovery from Hurricane Harvey (2017) in Houston, Texas, and applies a threshold-based spatial diffusion model to show how recovery in one neighborhood can catalyze or constrain recovery in adjacent areas. The concept of "recovery multipliers" emerges once again, but here it is applied at the neighborhood scale. Notably, the study finds that low-income and minority areas often serve as crucial multipliers, meaning that targeting these areas with resources can accelerate the overall recovery process more efficiently than if aid were distributed uniformly. Moreover, the study highlights heterogeneous spatial effects: some areas are well-connected, and their recovery quickly diffuses outward, while others are "recovery isolates" with slower or incomplete recovery that drags down the regional average. In practical terms, these results underscore that coordinated interventions across interdependent neighborhoods can produce far better outcomes than piecemeal strategies.

The study revealed stark heterogeneity in spatial network effects on recovery. Some areas benefited from strong positive spillover—being surrounded by recovering neighbors helped pull them up faster—while others did not. They observed the formation of recovery isolates: slow recovery hotspots which are clusters of vulnerable neighborhoods that lagged in recovery. For instance, a CBG "surrounded by other socially vulnerable areas and [lacking] spatial effect from neighboring areas" tended to struggle longer and could become an isolated pocket of slow recovery. This finding identified a mechanism explaining why certain low-resource communities experience prolonged recovery: their immediate neighbors are similarly resource-constrained and cannot provide the boost that more resilient neighbors might.

An important theoretical contribution arising from this work is the introduction of recovery multipliers. These are communities or nodes in the network whose recovery has a disproportionate positive influence on the overall system. The study showed that if socially vulnerable areas that act as recovery multipliers are identified and prioritized for aid, the entire community's recovery can be expedited "while also enhancing equity." In practice, this means strategically investing in certain key neighborhoods yields a multiplier effect, catalyzing broader recovery that cascades through the spatial network. This concept parallels the idea of influential spreaders in information networks.

This work crystallizes the argument that recovery is not an isolated phenomenon—it is an interdependent network event shaped by spatial adjacency, resource flow, and social-economic context. The identification of neighborhood-level multipliers reinforces the earlier concept of critical nodes or hubs. In line with the broader thesis, the study suggests that tackling inequality—by focusing on vulnerable or marginalized neighborhoods—can be both equitable and strategically optimal in terms of speeding system-wide recovery. By employing real-world data from a major U.S. metropolitan area, this study exemplifies the power of data-driven, network-diffusion modeling in guiding policymakers toward more effective, evidence-based disaster recovery strategies.

## 3   Concluding remarks

Across these six studies, a unifying theme emerges: networked processes fundamentally govern how communities experience, respond to, and ultimately recover from disasters, regardless of whether the focal network is social, economic, or spatial. Social networks underpin collective sense-making, where influential communicators and localized interactions guide how risk information diffuses and shapes public perception. In behavior-dependent business networks, interdependencies among establishments mean that swift reopening of certain "recovery multiplier" businesses can catalyze rebound across the entire economic ecosystem. Meanwhile, human mobility networks reveal that daily travel patterns determine "life activity flood exposure," extending flood risk far beyond a static notion of residents living in a flood zone. Turning to socio-spatial ties, hazard-exposure heterophily reveals how connections across risk boundaries foster the exchange of critical resources, reducing vulnerability in communities otherwise isolated by income and geography. Finally, spatial interdependencies at the neighborhood scale show that coordinated recovery efforts in low-income or minority "multiplier" areas can accelerate community-wide restoration through network diffusion effects. Taken together, these studies confirm that network structures and diffusion processes—ranging from communication channels to business dependencies and spatial adjacency—amplify or attenuate disaster impacts and recovery, underscoring the importance of harnessing these network dynamics to build more resilient and equitable communities.

Also, a network dynamics lens reveals a series of novel concepts and characterizations that explain how community resilience and recovery unfold beyond static assumptions of hazard and vulnerability. Hazard-exposure heterophily captures the degree to which people in at-risk areas connect with those in safer zones, influencing resource flows and overall recovery capacity. This notion intersects with emergent social cohesion, whereby weak ties that form abruptly during disasters serve as critical conduits for rapid information sharing and support. Similarly, latent hazard exposure emphasizes how daily mobility networks disperse risk beyond designated flood zones, while recovery multipliers highlight key nodes—be they businesses or neighborhoods—whose rapid rebound catalyzes broader system-wide recovery. Conversely, recovery isolates face compounding setbacks due to sparse connections, requiring targeted interventions to break negative feedback loops. Finally, heterogeneous recovery thresholds point to diverse dependencies and tipping points across communities: some areas recover autonomously, whereas others need a critical mass of neighbors or businesses to reestablish baseline functioning. Taken together, these concepts underscore how dynamic processes—rather than static structures—shape who is exposed, who bounces back quickly, and how resource flows create cascading effects throughout interconnected socio-spatial networks.

Despite the recent attention directed to and progress on examining network dynamics processes that shape community resilience and recovery, much work remains to be done in this emerging field. Below are several avenues for advancing network-based disaster research. Although existing literature highlights the importance of either social, spatial, or economic networks, real-world communities are constituted by multi-layered and overlapping networks. For instance, an individual's flood exposure depends on mobility patterns (a spatial network), but their ability to receive assistance may hinge on social connections (a social network). Also, critical infrastructure—such as energy grids or communication systems—forms a separate physical network that can fail or be repaired, altering the broader connectivity structure. Future research should combine these different network layers into integrated models, capturing feedback loops where a failure in one network triggers cascading effects in others. This multi-layer approach could provide a more holistic, systems-level understanding of how communities respond to and recover from disasters. Many of the studies adopt or adapt epidemiological or threshold-based diffusion frameworks to model how recovery or hazard exposure propagates. However, deeper insight is needed into the mechanisms that underlie these diffusion processes. For example, how do trust, social norms, and collective efficacy influence whether information or assistance flows freely among neighbors? When do communication bottlenecks arise, and how can they be alleviated? Are there specific interventions—like placing official messages in the hands of known influencers or subsidizing key businesses—that can reliably "activate" beneficial network cascades? Investigation into these causal mechanisms, grounded in social theory and tested with empirical data, can refine the predictive power of diffusion models and guide policy levers for improved disaster management. Disasters not only exacerbate existing vulnerabilities in network structures but also reshape them. Roads are rebuilt, businesses relocate or permanently close, and individuals may migrate, forming new social ties in different places. Over multiple events, these adaptive or reactive changes can accumulate, transforming the long-term resilience of a region. There is a pressing need for longitudinal research that tracks how networks evolve post-disaster and what that implies for future hazard preparedness. Such studies could uncover, for instance, whether repeated floods drive residents to move away, fragmenting local social networks, or whether certain communities grow more cohesive through shared coping experiences. Understanding this evolution would allow policymakers to anticipate new forms of vulnerability (or emergent strengths) that arise over time.

All six studies highlight the interplay between network dynamics and inequality. Low-income neighborhoods, minority communities, and underserved populations consistently face higher exposure, slower recovery, or both—often due to network constraints such as limited mobility, fewer bridging social ties, or inadequate infrastructure. Research should move beyond simply documenting these disparities to designing and testing equitable interventions. For instance, can targeted grants to "recovery multiplier" neighborhoods deliver faster and more equitable outcomes for the entire city? Can expansions of public transit reduce the mobility-based risk discovered in LAFE metrics? Can digital connectivity initiatives foster heterophily across hazard boundaries, thereby closing resource gaps? By systematically evaluating the distributional impacts of network-based policies, scholars and practitioners can push resilience research toward more justice-oriented paradigms. The explosion of large-scale data from smartphones, social media, and other digital platforms has led to new empirical insights, but theoretical integration lags behind. Much of the work so far has focused on identifying patterns and correlations, leaving open questions about generalizability and causal inference. Future research must engage more explicitly with theoretical frameworks in disaster science, sociology, epidemiology, urban planning, and related fields to ensure that big data findings inform—and are informed by—broader conceptual constructs. By grounding computational models in robust theory, the field can move toward predictive and prescriptive capabilities that not only explain observed patterns but also guide decision-making for more resilient, equitable communities.

In sum, the emerging scholarship on network dynamics in community resilience and recovery calls for a transformative shift in how we conceptualize and address the impacts of disasters. From social media–mediated risk perception to business interdependencies, from person-centric flood exposure metrics to hazard-exposure heterophily, and from spatial diffusion models identifying recovery multipliers to real-world data on mobility and social connectedness—each line of inquiry converges on the insight that networks are fundamental to understanding disaster outcomes. By elucidating how diffusion processes, inequalities, and critical nodes intersect, network-based disaster research offers a powerful analytical lens for diagnosing vulnerabilities, designing targeted interventions, and fostering collective resilience. Crucially, these approaches also open the door to interdisciplinary collaborations, linking computational modeling with social theory, urban planning, public policy, and emergency management. Such cross-cutting efforts are poised to reshape not only academic discourse but also the real-world strategies used by governments, NGOs, and communities as they grapple with ever more complex hazard landscapes. Looking ahead, the integration of multi-layer network analyses, mechanistic diffusion studies, longitudinal network evolution research, equity-focused interventions, and big data–informed theory building will be pivotal. By embracing the network paradigm, researchers and practitioners can devise more nuanced, effective, and just responses to the crises that threaten today's interconnected world. The ultimate aspiration is clear: to harness the very forces of connectivity, interdependence, and collective action for the good of all, ensuring that the communities of tomorrow can withstand, adapt, and recover from even the most daunting adversities.

## Acknowledgments

This material is based in part upon work supported by the National Science Foundation under Grant CMMI-1846069 (CAREER). Any opinions, findings, conclusions, or recommendations expressed in this material are those of the authors and do not necessarily reflect the views of the National Science Foundation.

## Author contributions

All authors critically revised the manuscript, provided final approval for publication, and agreed to be held accountable for the work performed therein. C.F., the Ph.D. student researcher, conceived the idea, collected the materials, and drafted the manuscript. A.M., the faculty advisor, offered critical feedback on the project's development and manuscript.

## Competing interests

The authors declare no conflict of interest.